# The enabling electronic motif for topological insulation in $ABO_3$ perovskites and its structural stability

Xiuwen Zhang[1,2,*], Leonardo B. Abdalla[1], Qihang Liu[1,*], and Alex Zunger[1,*]
[1]Renewable and Sustainable Energy Institute, University of Colorado, Boulder, Colorado 80309, USA
[2]College of Electronic Science and Technology, Shenzhen University, Shenzhen, Guangdong, P. R. China

*Email: xiuwen.zhang@colorado.edu; qihang.liu@colorado.edu; alex.zunger@colorado.edu

## Abstract

Stable oxide topological insulators (TI's) that could bring together the traditional oxide functionalities with the dissipationless surface states of TI's have been sought for years but none was found. Yet, heavier chalcogenides (selenides, tellurides) were readily found to be TI's. We clarify here the basic contradiction between TI-ness and stability which is maximal for oxides, and trace the basic design principles necessary to identify the window of opportunity of stable TI's. We first identify the electronic motif that can achieve topological band inversion ('topological gene') in $ABO_3$ as being a lone-pair electron-rich B atom (e.g. Te, I, Bi) at the octahedral site. We then illustrate that poorly screened oxide systems with large inversion energies can undergo energy-lowering atomic distortions that remove the band inversion. We identify the coexistence windows of TI functionality and structure stability for different pressures and find that the common cubic $ABO_3$ structures have inversion energies lying outside this coexistence window at zero pressure but could be moved into the coexistence window at moderate pressures. Our study demonstrates the interplay between topological band inversion and structural stability and traces the principles needed to design stable oxide topological insulators at ambient pressures.

## 1. Introduction

Topological insulators (TI's) are materials having an inverted order of the occupied valence and unoccupied conduction bands at time-reversal invariant (TRI) wave vectors in the Brillouin zone (BZ), and can be characterized by the topological invariant[1] $Z_2 = 1$. Theory then assures that in lower-dimensional forms (2D surface or 1D edge) of the topological bulk system there will be states that possess passivation-resistant, linearly dispersed and mutually crossing (metallic) energy bands[2]. The required band inversion in the parent bulk system is generally achieved by introducing high atomic number (Z) cations and anions having strong spin-orbit coupling (SOC)[3-8]. However, such heavy-atom compounds tend to pose defective crystal structures (e.g., spontaneous vacancy formation causing metallic, not insulating behavior) associated with the relatively low cohesion of the weak heavy-atom chemical bonds[9-11]. The recent quest of topological insulators in oxides[12-18] has been partially motivated by the hope that this will deliver defect-tolerant lattices, often characteristic of metal oxides[19], while at the same time affording the integration of topological properties with the rich oxide functionalities such as transparent conductivity[20], ferroelectricity[21], ferromagnetism[22], or superconductivity[23]. However, the electronic structures of common octet metal oxides, such as $ABO_3$ perovskites or $A_2BO_4$ spinels, show that while they may be stable and have wide energy band gaps[19] $E_g$, they generally lack band inversion, having a trivial topological invariant $Z_2 = 0$. For example, the common $ABO_3$ perovskite have oxygen-derived valence bands and B-atom-derived conduction bands [see **Figure 1**(a,b)], hence no band inversion. Indeed, whereas numerous topological insulators have been experimentally observed in selenides and tellurides (e.g. $Bi_2Se_3$ [4] and $Bi_2Te_3$ [24], stable oxide TI's are yet unknown. Anecdotal examples (discussed in Supplementary Section I) abound of theoretically proposed wide gap oxide TI's in *assumed hypothetical* crystal structures that turned out, however, to be significantly unstable when energy-lowering structural relaxation were explored in such hypothetical structures[12-18, 25-30].

We conjecture that the conditions needed for TI-ness—depopulation of bonding valence band states and the occupation of anti-bonding conduction band states—may be contraindicated to thermodynamic stability, if carried out throughout a significant portion of the Brillouin zone. If left unscreened (as in strongly ionic systems), such destabilizing forces may drive structural deformations that alter the crystal symmetry and could undo the band inversion, as illustrated below. Although metastable structures can certainly be made[31-33], it would be desirable to predict compounds that are TI in not-too-unstable structures, which can be synthesized without fear of producing a topologically trivial but stabler structure, or decomposing after synthesis to a combination of phases that may not be TI's. In this work we report the results of *ab-initio* co-evaluation of TI-ness and stability for a class of oxides perovskites, elucidating the physical origin of the hitherto mysterious difficulty to obtain simultaneously the *electronic* features leading to TI-ness ('topological gene') and the *structural* features of the corresponding lattice structure in $ABO_3$ oxides. We identify the topological gene that can achieve topological band inversion in oxide perovskites as being a lone-pair electron-rich B atom (e.g., Te, I, Bi) rather than Ti, Nb, Y, respectively) at the octahedral site in the cubic $ABO_3$. Oxides tend to have larger predicted inversion energies ($\Delta_i$ > 1 eV) than selenides or tellurides ($\Delta_i$ < 0.4 eV[4, 24]), and are generally less covalent than the heavier chalcogenides where lattice distortions would be partially screened by the more delocalized electronic states. We then illustrate that poorly screened oxide systems with large inversion energies can undergo energy-lowering atomic distortions that remove the topological band inversion. Favorable screening exists in oxides at higher pressures, and lead to stable oxide TI's as illustrated here for $BaTeO_3$ and $RbIO_3$ under moderate pressures. *Covalent oxides* with strong metal-oxygen orbital mixing[34] do exist and point to the direction of future search for stable oxide TI's at ambient pressures.

## 2. Topological gene and stability gene

To address this issue we introduce two constructs: We first identify an *electronic* motif within a group of $ABO_3$ oxides that would generate band inversion—the *"topological gene"* of this group of compounds. As Fig. 1(c, d) illustrates the "topological gene" here is the octahedral $BO_6$ motif with *lone pair B atom* (generated, e.g., by replacing an electron-poor Ti atom in $BaTiO_3$ by an electron-rich Te atom in $BaTeO_3$ that has an additional $d^{10}s^2$ shell). Second, we examine the stability of the crystal structure that hosts the topological gene, relative to the ground state structures that hosts the "*stability gene*" of this group of compounds. The key challenge is to see if structures with the topological gene can also have the stability gene. Thus, *co-evaluation* of the electronic structure and stability is required.

We confirm via calculated band structures in the framework of density function theory (DFT) using projector-augmented wave (PAW) pseudopotential with the exchange-correlation of Perdew-Burke-Ernzerhof (PBE)[35-37], and the calculated topological invariant[38, 39] $Z_2$ (see Methods Section and Supplementary Section II for the details of calculation methods) that twelve $ABO_3$ compounds ($BaTeO_3$, $SrTeO_3$, $CaTeO_3$, $BaSeO_3$, $SrSeO_3$, $CaSeO_3$, $RbIO_3$, $KIO_3$, $NaIO_3$, $RbBrO_3$, $KBrO_3$, $NaBrO_3$) in the *assumed* cubic perovskite structure with lone-pair B atoms at the octahedral site are in fact TI's, whereas the corresponding compounds with non-lone-pair B ions at the octahedral site (e.g., $BaTiO_3$ and $KNbO_3$) are found to be normal insulators. This substantiates the identity of the topological gene in $ABO_3$ compounds. However, the crystal structure that host this topological gene is not the stablest structure at this composition: total energy relaxation calculations reveal that the cubic topological structures would relax to lower energy non-cubic perovskite phases that are not TI's. This is consistent with the above-mentioned topology *vs* stability conjecture. Indeed, by performing constrained DFT calculations (see Methods Section), we observe a regained stability of the cubic phase. We further identified the window of coexistence of TI-ness and stability at ambient conditions, and find that the topological cubic $ABO_3$ phases have inversion energies lying outside this coexistence window. Building upon this identification, we find that moderate external pressure can significantly expand the above coexistence window, and eventually leads to simultaneously stable and topological cubic $ABO_3$ phases, thus bringing the 'stability gene' into coincidence with the 'topological gene'. This study illustrates the interplay between the topological gene needed to obtain band inversion and the structural motif needed for stability in the $ABO_3$ oxides, and traces the window of opportunity to design stable oxide TI's at ambient pressures.

### 3. Electronic requirements for the topological gene in cubic $ABO_3$ perovskites

In conventional $ABO_3$ compounds with electron-poor B atoms [such as illustrated in Fig. 1(b)] the occupied valence band is oxygen derived whereas the empty conduction states are B-atom derived with normal, *s*-below-*p* orbital order (thus, no inversion). To induce band inversion, one would like to change the order of B atom orbitals to *p*-below-*s* and assure that the Fermi energy is located between these occupied and unoccupied states, respectively. The idea is to replace the B atom in $ABO_3$ in the cubic perovskite structure [see the inset of Fig. 1(c)] by an electron-rich B element of the same formal charge. Such a B atom has an occupied (lone pair) *s* orbital below its empty *p* orbital, yet above the O-*p* state, as shown in Fig. 1(c). The lone pair *s* orbital can thus become an unoccupied state above the conduction band due to (B-atom *s*)–(oxygen *p*) level repulsion, leading to band inversion (B-*p* below B-*s*) as well as to a finite excitation band gap between B-*p* (occupied) and B-*s* (unoccupied), which is illustrated in Fig. 1(d). We will illustrate via quantitative DFT calculations the above concept and its rather broad applicability for different groups of perovskite compounds.

In the **$II_A$-$IV_A$-$O_3$** group of compounds exemplified by $BaTiO_3$, we replace $Ti^{4+}$ by the electron-richer $Te^{4+}$ having additional $d^{10}s^2$ shell, leading to the **$II_A$-$VI_B$-$O_3$** group of compounds exemplified by $BaTeO_3$. Two effects, noted by our detailed DFT calculations and illustrated by a simple orbital diagram [see Fig. 1] are associated with this transmutation: (i) Because of the addition of $d^{10}s^2$ shell (where the $s^2$ lone-pair band is occupied), the outmost B-*d* and B-*s* states of B = Te become occupied, with a band gap located between B-*s* and B-*p* states [see Fig. 1(c,d)]. The outer *s, p,* and *d* atomic orbitals of B = Te drop in energy relative to B = Ti (because of the less screening of the core by the valence shell). (ii) The relativistic Darwin[40] effect causes the *s* orbital to become relatively localized and introduce an *occupied* lone-pair *s*-band, which in this case lies above the O-*p* bands. Since in the cubic perovskite structure Te (in $BaTeO_3$) or Ti (in $BaTiO_3$) are located at the octahedral $O_h$ site where they are bonded to six O atoms, there will be a level repulsion between the O-*p* and Te-*s* bands, displacing the Te-*s* band upwards. If the repulsion is strong enough, as it is at the R point in the BZ [see Fig. 1(e)], this repulsion will place Te-*s* *above* Te-*p* [see Fig. 1(d)], leading to band inversion between Te-*s* and Te-*p* states at *R* point in the Pm-3m structure [see **Figure 2**(a)]. *Therefore, the use at the B site of the electron-rich version (Te in place of Ti) causes band inversion, thus constituting the topological gene in these systems.*

Fig. 1(a,e) illustrates this design principle via DFT band structure. We see that while in $BaTiO_3$ the conduction band minimum (CBM) is composed of Ti-atom 3*d* states (yellow) throughout BZ, in $BaTeO_3$ there is a band inversion at R point where the valence band maximum (VBM) is made of Te-*p* orbital (green) instead of Te-*s* orbital (red). Calculation of the topological invariant from the wavefunctions (see Methods Section) finds $Z_2$ = 1 in $BaTeO_3$ as compared to the expected $Z_2$ = 0 in $BaTiO_3$. That this design principle is rather general can be seen by considering other leading $ABO_3$ perovskite prototypes.

In the **$I_A$-$V_A$-$O_3$** group of compounds exemplified by $KNbO_3$, we replace $Nb^{5+}$ by $I^{5+}$ leading to the group **$I_A$-$VII_B$-$O_3$** exemplified by $KIO_3$. Supplementary Section III compares the electronic structures of $KNbO_3$ and $KIO_3$. We find $Z_2$ = 0 in $KNbO_3$ whereas $Z_2$ = 1 in $KIO_3$ [41]. The Supplementary Section IV provides the values of the excitation gaps in the compounds designed according to the topological gene, and discusses effects due to possible DFT errors as compared to Hybrid functional (HSE06 [42]) calculations. The results show that the inversion energy ($\Delta_i$), i.e., the gap between the inverted valence and conduction bands at TRI wave vectors, is slightly reduced (by 4%) in HSE06 as compared to DFT, indicating that the band inversion discussed here is quite robust against DFT errors.

**4. Band inversion is often contraindicated to stability**

Having identified and verified an *electronic* motif—the topological gene—we next ask whether the crystal structure (here, cubic $ABO_3$ perovskite) that harbors the topology coincides with the structure that harbors stability.

**Figure 3** shows the total energies (in meV/atom, given in parentheses) of various $ABO_3$ structures (denoted S1-S15 in Fig. 2, S1 being the cubic perovskite structure while S2-S15 being the lowest-energy structures of specific $ABO_3$ compounds), relative to the lowest-energy phase. In addition to stability, Fig. 3 also denotes if according to the calculated topological invariant $Z_2$ the compound is a TI or a normal insulator (NI). *We find from such total energy minimizations that the $BO_6$ octahedral unit with the said electron rich B atom tends to distort towards a stabler, non-cubic crystal structures [29, 30] and that this distortion removes the band inversion and thus TI-ness*. Therefore, the stability gene and TI gene tend to contradict each other for the $ABO_3$ compounds at ambient conditions.

To get a deeper understanding on the interplay between structural stability and band inversion, we perform constrained DFT calculations constructed for tuning band inversion to examine its effect on the total energy. The tuning of the inversion energy can be done by using an external potential that shifts upwards the B atom *p* orbital energy, thus, according to Fig. 1(d) the inverted structure (*p*-below-*s*) can be tuned to be un-inverted (*s*-below-*p*). This constraint can be implemented, for example, by adding an external potential term[43] $V_p$ to the DFT Hamiltonian acting on the I-*p* orbital in $RbIO_3$. We then monitor the total energy of the cubic perovskite structure (S1) relative to its stable R3m rhombohedral phase [S2, see Fig. 2(b)], as a function of $\Delta_i$. **Figure 4**(a) shows that as $\Delta_i$ decreases, the energy of the cubic S1 phase (relative to S2) also decreases, indicating that band inversion is contraindicated with the stability of crystal structure (see Supplementary Section V for the evolution of the electronic structures with decreasing inversion energy). *Significantly, however, there is a narrow window (0 < $\Delta_i$ < 0.8 eV) shown by red background in Fig. 4(a), where topological band inversion and the stability of the band-inverted crystal structure can coexist.* It is noticed that the currently known stable TI's (e.g. $Bi_2Se_3$ [4] and $Bi_2Te_3$ [24]) have $\Delta_i$ much smaller than 0.8 eV, lying inside the coexistence window (0 < $\Delta_i$ < 0.8 eV). Whereas the cubic $ABO_3$ structures with short B-O bonds have $\Delta_i$ > 1 eV due to the strong coupling between B-*s* and O-*p* states, lying outside of the coexistence window. Instead of searching for stable oxide TI's with small $\Delta_i$, we attempt to stabilize oxide TI's with large inversion energies.

## 5. Bringing the topological gene into overlap with the stability gene by applying pressure

Inspired by the fact that many $ABO_3$ compounds (e.g. $BaTiO_3$ [44]) that are not cubic at ambient conditions can be stabilized under external pressure in the cubic form, we test the effect of external pressure on the interplay between band inversion and structural stability. We perform similar calculations as Fig. 4(a), but apply external hydrostatic pressure of ≤ 20 GPa. The result of Fig. 4(b) show that the coexistence window for band inversion and structural stability (red background) is significantly extended. At 20 GPa, a stable TI structure can form while tolerating a band inversion of the magnitude of about 2.5 eV. To demonstrate the dramatic effect of pressure on TI-ness versus stability diagram [Fig. 4(b)], we perform enthalpy calculations for $BaTeO_3$ (from group $II_A$-$VI_B$-$O_3$) and $RbIO_3$ (from group $I_A$-$VII_B$-$O_3$). **Figure 5**(a,b) shows the calculated enthalpies of the relevant structure types for $BaTeO_3$ and $RbIO_3$, as a function of hydrostatic pressure, which demonstrate that the cubic perovskite [S1, see red squares in Fig. 5(a,b)] tends to be stabilized by external pressure. At pressure of 15 GPa (35 GPa), the S1 phase that contains the topological gene becomes the lowest-enthalpy structure for $BaTeO_3$ ($RbIO_3$). We further check the effect of external pressure on the band inversions in cubic $BaTeO_3$ and $RbIO_3$, finding that the band inversion is not removed by external pressure [see Fig. 5(c,d)], but on the contrary, increased by the pressure from 2.3 eV at 0 GPa to 3.4 eV at 35 GPa for $BaTeO_3$, and from 3.0 eV at 0 GPa to 3.8 eV at 35 GPa for $RbIO_3$. It is expected because the pressure decreases the atomic bond lengths of $BO_6$ octahedral, leading to larger *s*-*p* repulsion. The enhanced tolerance of cubic $ABO_3$ structures to band inversions is because of the increased electronic screening effect under pressures. Similar to heavier-chalcogenide TI's (e.g. $Bi_2Se_3$), the lattice distortions would be screened by the delocalized electronic states. Here, we use external pressure (as an example of external constraints) to tune the screening effect in oxides. Strong screening could exist in oxides without external constraints, such as *covalent oxides* with strong metal-oxygen orbital mixing[34], pointing to the direction of future search of stable oxide TI's with large inversion energies that is robust under perturbations such as thermal expansion.

## 6. Conclusions and discussion

Studying the interplay between TI-ness and stability in search of oxide topological insulators opens the door of several scenarios: (i) understanding the general conflicting trends between TI-ness and stability answers the question concerning the rareness of experimentally realizable topological phases, and hopefully discourage false positive predictions of topological insulators in hypothetical, highly unstable structures. (ii) These trends can be readily applied to other band inversion quantum phases, such as topological Dirac semimetals and Weyl semimetals. (iii) The results in this study suggest the directions for future search of stable oxides TI's, that are (a) search for oxides with small inversion energies (b) design large inversion-energy oxide TIs that can be stabilized by external constraints or intrinsic screening effect. This opens the way of rational design of stable oxide TI's using topological genes as pivots, and electronic screening/external constraints (e.g. pressure and strain) as levers.

## Methods

***DFT total energy and electronic structure evaluation.*** Total energies were calculated using the projector-augmented wave (PAW) pseudopotential[37] total energy method without spin-orbit coupling with the exchange-correlation of Perdew-Burke-Ernzerhof (PBE)[35] as implemented in the Vienna ab-initio simulation package (VASP)[36]. We use an energy-cutoff of 520 eV and reciprocal space grids with densities of 2π×0.068 $Å^{-1}$ and 2π×0.051 $Å^{-1}$ for relaxation and static calculation, respectively. Electronic structure and physical properties were calculated from DFT taking into account spin-orbit coupling by a perturbation $\sum_{i,l,m} V_l^{SO} \vec{L} \cdot \vec{S} |l,m\rangle_{i\,i}\langle l,m|$ to the pseudopotential[45] ($|l,m\rangle_i$ is the angular momentum eigenstate of atomic site $i$). To illustrate the band inversions, we project the calculated wave function $|\phi\rangle$ on spin and orbital basis of each atomic site $C_{i,l,m,\eta} =$ $\langle\phi|(s_\eta \otimes |l,m\rangle_{i\,i}\langle l,m|)|\phi\rangle$ and then sum $C_{i,l,m,\eta}$ for a given atomic orbital. The Wigner-Seitz radii for constructing $|l,m\rangle_i$ used in this study are listed in the pseudopotentials of the VASP simulation package[36]. To illustrate the interplay between band inversion and structural stability, we performed constrained DFT calculations by applying a Hubbard-like external potential term[43] ($V_p$) on the I-*p* orbital in $RbIO_3$ in the Hamiltonian to gradually remove the band inversion between I-*p* and I-*s* states.

***Z₂ characterization.*** For centrosymmetric structures, we evaluate $Z_2$ from the parities of the wave functions at the time reversal invariant *k*-points[39]. For non-centrosymmetric structures, we use the method from Refs.[38, 46, 47] to calculate the topological invariant $Z_2$, which is based on the evolution of Wannier function centers. The topological invariant $Z_2$ is expressed as the number of times mod 2 of the partner switching between the Wannier function centers during the evolution (as illustrated in Fig. S3). To directly look at the evolution of Wannier function centers, we follow the scheme proposed by Yu *et al.*[38] The topological nature of a 3D compound is determined by looking at two effective 2D systems with $k_z = 0$ and $k_z = \pi$. Each 2D effective system is then considered as a series of effective 1D system with fixed $k_y$. For an effective 1D system with periodic condition, the unitary position operator is defined as,

$$\hat{x} = \sum_{j\alpha} e^{-i\delta_{k_x}\cdot \mathbf{R}_j} |j\alpha\rangle\langle j\alpha|, \quad (1)$$

where $\delta_{k_x} = \frac{2\pi}{N_x a_x}$, $N_x$ is the number of real space unit cells along $x$ direction, $\mathbf{R}_j$ denotes the position of the $j$th lattice site, and $\alpha$ denotes the index of the Wannier function $|j\alpha\rangle$. The projection operator for the occupied subspace for a fixed $k_y$ can be defined as,

$$\hat{P}_{k_y} = \sum_{n\leq 2N,\, k_x} |\psi_{n\mathbf{k}}\rangle\langle\psi_{n\mathbf{k}}| \ ; \ \text{for } k_z = 0 \text{ or } k_z = \pi, \quad (2)$$

where $|\psi_{n\mathbf{k}}\rangle$ is the Bloch state and $2N$ is the number of occupied states.

We then consider the eigenvalue of the projected position operator,

$$\hat{x}_P(k_y) = \hat{P}_{k_y} \hat{x} \hat{P}_{k_y}, \tag{3}$$

that has the following form:

$$\hat{x}_P(k_y) = \begin{bmatrix} 0 & X_{0,1} & 0 & 0 & \cdots & 0 \\ 0 & 0 & X_{1,2} & 0 & \cdots & 0 \\ 0 & 0 & 0 & X_{2,3} & \cdots & 0 \\ \vdots & \vdots & \vdots & \vdots & \vdots & \vdots \\ 0 & 0 & 0 & 0 & \cdots & X_{N_x-2,N_x-1} \\ X_{N_x-1,0} & 0 & 0 & 0 & \cdots & 0 \end{bmatrix}, \tag{4}$$

where $X_{0,1}$, $X_{1,2}$, ... are $2N\times2N$ matrices. The eigenvalue of the projected position operator can be solved by the transfer matrix method. We define a product of $X_{0,1}$, $X_{1,2}$, ... as,

$$D(k_y) = X_{0,1}X_{1,2}X_{2,3}\cdots X_{Nx-2,Nx-1}X_{Nx-1,0}. \tag{5}$$

$D(k_y)$ has $2N$ eigenvalues:

$$\lambda_m^D(k_y) = |\lambda_m^D(k_y)|e^{i\theta_m^D(k_y)}, m = 1, \ldots 2N. \tag{6}$$

The evolution of the Wannier function center with $k_y$ can be easily obtained by looking at the phase factor $\theta_m^D$ (named Wannier center for convenience). The Wannier centers are paired at $k_y = 0$ and $k_y = \pi$, and during the evolution from $k_y = 0$ to $k_y = \pi$, the Wannier center can switch partners, with an odd number of partner switching corresponding to $Z_2 = 1$ (see Supplementary Section II).

## Supporting Information

Supporting Information is available from the Wiley Online Library or from the author.

## Acknowledgements

Part of the work on calculation of topological properties was supported by NSF Grant titled "Theory-Guided Experimental Search of Designed Topological Insulators and Band-Inverted Insulators" (No. DMREF-1334170). Work on calculation of material stability was supported by Office of Science, Basic Energy Science, MSE division under grant DE-SC0010467 to CU Boulder. This work used the Extreme Science and Engineering Discovery Environment (XSEDE), which is supported by NSF grant number ACI-1053575. We thank Dr. Saicharan Aswartham and Prof. Gang Cao for sharing with us the crystallographic information of the newly synthesized and characterized $BaTeO_3$ ($P2_1/c$) structure, and thank Prof. Dan Dessau for helpful discussions.

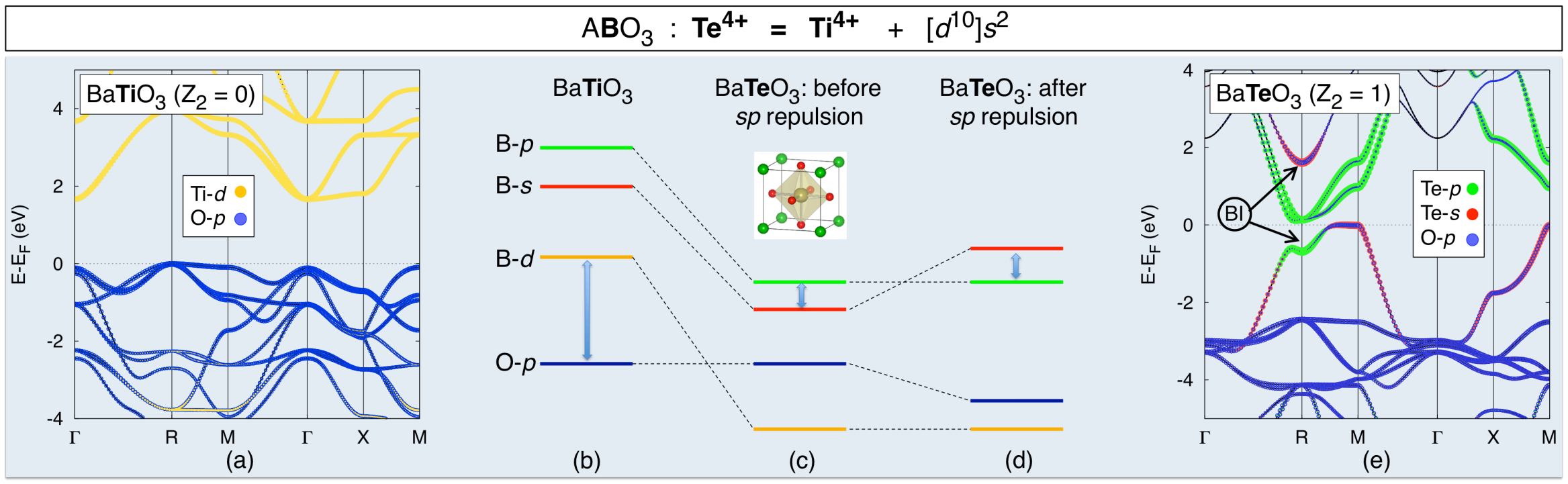

FIG. 1. Illustration of the topological gene in cubic $ABO_3$ by electronic structures in DFT of $BaTiO_3$ (a) and $BaTeO_3$ (e) with cubic perovskite (Pm-3m) structure (at zero pressure), and schematic orbital diagram of $ABO_3$ perovskite compounds, in the cases of (b) electron-poor B atom such as Ti in $BaTiO_3$ where there is no band inversion, (c) and (d) electron-rich ("lone pair") B atom such as Te having additional $d^{10}s^2$ orbital shells, leading to band inversion in $BaTeO_3$. The vertical blue arrows indicate the fundamental band gaps. The inset in (c) shows the "topological gene"—associated with topological insulation in this class of compounds. BI denotes band inversion with arrows pointing to the inverted states. The dotted lines with different colors denote the band projection onto different atomic orbitals.

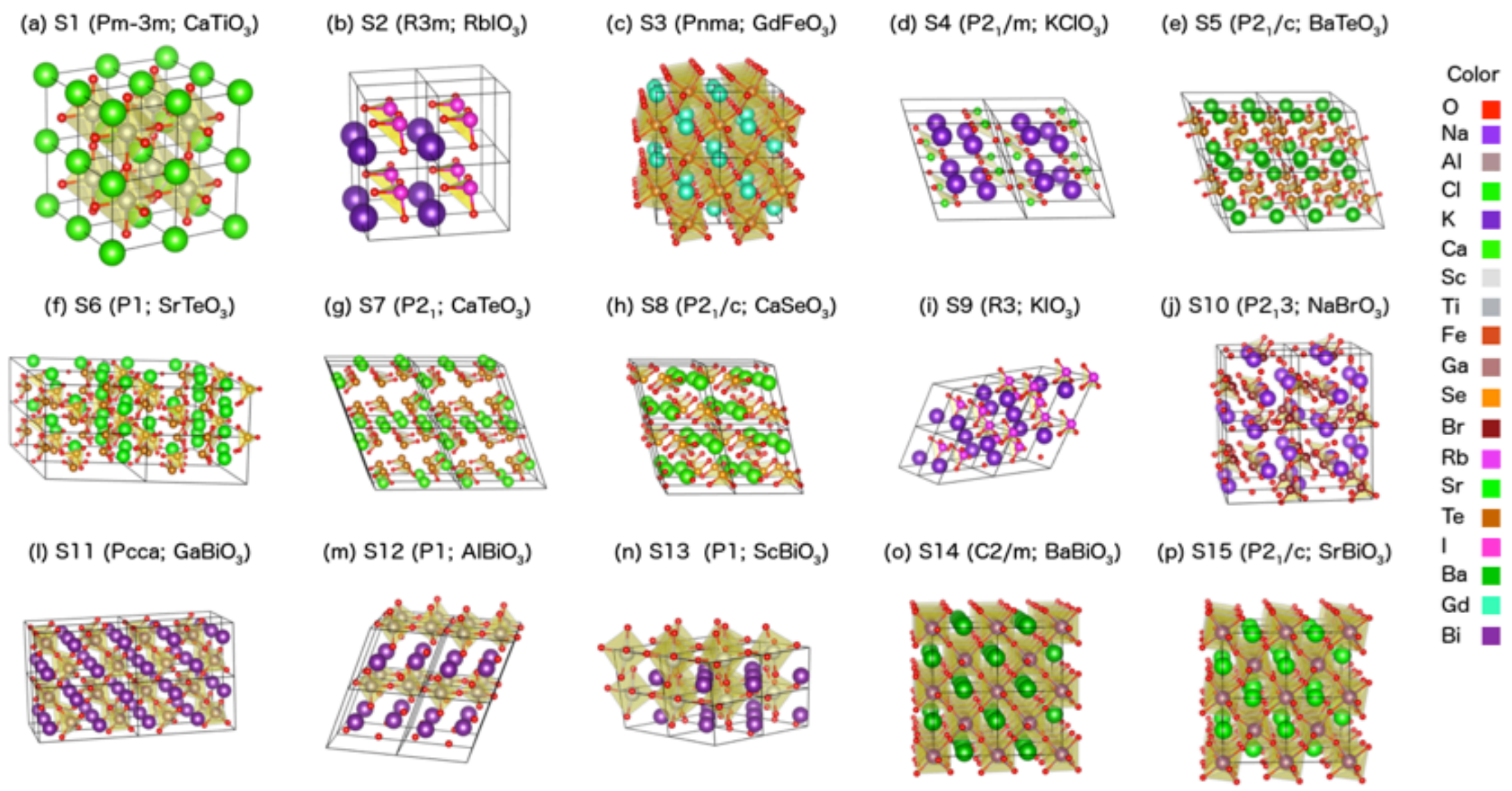

FIG. 2. Illustration of the geometry of the structures of $ABO_3$ discussed (a-p). The legend of coloring for distinguishing atoms of different chemical elements is displayed on the right. The black boxes in each structure represent the unit cell, and 2×2×2 supercells were shown for better visual understanding. All triangle or tetragonal yellow structure corresponds to bonding from elements to oxygen with a distance lower than 2.4 Å.

| $II_A$ \ $VI_B$ | Te | Se |
|---|---|---|
| Ba | $BaTeO_3$<br>S1 (156) : $R^1$<br>S2 (76) : NI<br>S3 (37) : NI<br>S4 (3) : NI<br>S5 (0) : NI | $BaSeO_3$<br>S1 (366) : $R^1$<br>S2 (75) : NI<br>S3 (63) : NI<br>S4 (0) : NI |
| Sr | $SrTeO_3$<br>S1 (232) : $R^1$<br>S2 (43) : NI<br>S4 (14) : NI<br>S6 (9) : NI<br>S3 (0) : NI | $SrSeO_3$<br>S1 (353) : $R^1$<br>S2 (68) : NI<br>S3 (25) : NI<br>S4 (0) : NI |
| Ca | $CaTeO_3$<br>S1 (426) : $R^1$<br>S4 (76) : NI<br>S2 (52) : NI<br>S3 (17) : NI<br>S7 (0) : NI | $CaSeO_3$<br>S1 (449) : $R^1$<br>S2 (46) : NI<br>S4 (46) : NI<br>S3 (12) : NI<br>S8 (0) : NI |

| $I_A$ \ $VII_B$ | I | Br |
|---|---|---|
| Rb | $RbIO_3$<br>S1 (241) : $R^1$<br>S3 (37) : NI<br>S4 (24) : NI<br>S2 (0) : NI | $RbBrO_3$<br>S1 (551) : $R^1$<br>S3 (20) : NI<br>S4 (5) : NI<br>S2 (0) : NI |
| K | $KIO_3$<br>S1 (190) : $R^1$<br>S3 (22) : NI<br>S4 (21) : NI<br>S2 (6) : NI<br>S9 (0) : NI | $KBrO_3$<br>S1 (476) : $R^1$<br>S3 (9) : NI<br>S2 (3) : NI<br>S4 (0) : NI |
| Na | $NaIO_3$<br>S1 (224) : $R^1$<br>S2 (56) : NI<br>S4 (34) : NI<br>S3 (0) : NI | $NaBrO_3$<br>S1 (469) : $R^1$<br>S2 (51) : NI<br>S4 (36) : NI<br>S3 (7) : NI<br>S10 (0) : NI |

| G. \ A | Ga/Y/Ba | Al/Sc/Sr |
|---|---|---|
| $III_B$ | *$GaBiO_3$*<br>S1 (223) : NI<br>S2 (76) : NI<br>S3 (71) : NI<br>S4 (48) : NI<br>S11 (0) : NI | *$AlBiO_3$*<br>S1 (97) : NI<br>S4 (81) : NI<br>S3 (44) : NI<br>S2 (38) : NI<br>S12 (0) : NI |
| $III_A$ | *$YBiO_3$*<br>S1 (629) : NI<br>S4 (39) : NI<br>S2 (31) : NI<br>S3 (0) : NI | *$ScBiO_3$*<br>S1 (325) : NI<br>S4 (77) : NI<br>S2 (27) : NI<br>S3 (10) : NI<br>S13 (0) : NI |
| $II_A$ | $BaBiO_3$<br>S4 (76) : NI<br>S1(23): $R^1$ e<br>S3 (4) : NI<br>S14 (0) : NI | $SrBiO_3$<br>S1(142):$R^1$ e<br>S2 (119) : NI<br>S4 (69) : NI<br>S3 (14) : NI<br>S15 (0) : NI |

FIG. 3. Presence of topological properties (indicated by red, with $R^1$ denoting the single band inversion at R point in BZ) *vs* absence (indicated by black, with NI denoting normal insulator) and relative DFT total energies (meV/atom, with zero indicating the ground state, shown in parentheses) of $ABO_3$ compounds with different crystal structures S1-S15 (see Fig. 2) at zero pressure. The calculations included swapping of A and B elements on the $ABO_3$ atomic sites and the lower-energy configuration is selected. All considered compounds except $YBiO_3$, $ScBiO_3$, $GaBiO_3$ and $AlBiO_3$ (shown in italics; note that in their lower-energy S1 structure, Bi is at the A site), are thermodynamically stable in their lowest-energy structure, as found by DFT calculations of the convex hull [29, 30]. For $BaBiO_3$, the initial S2 structure relaxes into the S1 structure, thus the total energy of S1 but not S2 structure is reported. For $BaBiO_3$ (as well as $SrBiO_3$) in the S1 structure, the band inversion (denoted as $R^1$e) occurs only after doping by 1 electron/formula unit (see Supplementary Fig. S1), but not at the band edges as for the other cases.

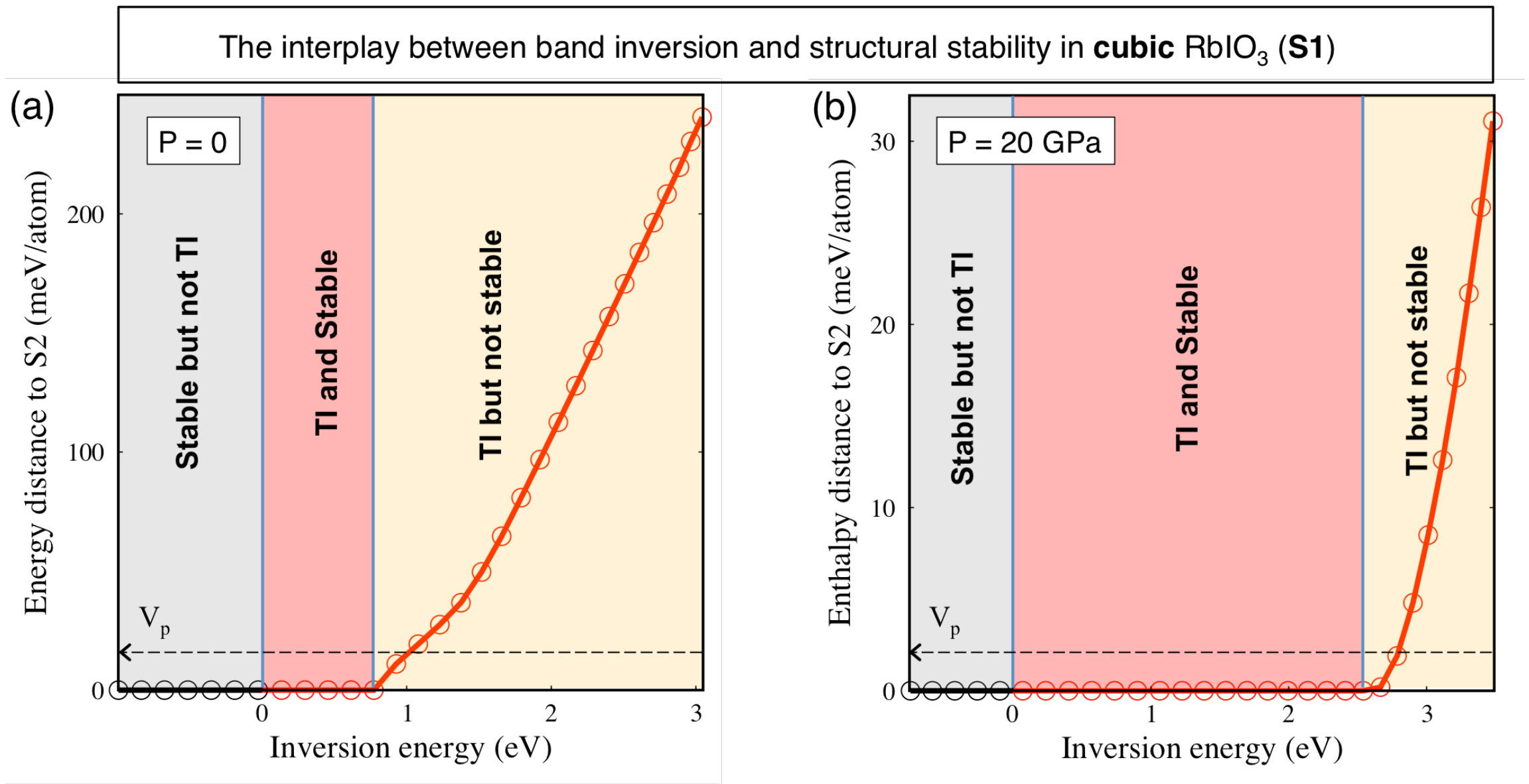

FIG. 4. Illustration of the interplay between band inversion and structural stability in cubic $RbIO_3$ at 0 GPa (a) and 20 GPa (b). The inversion energy in the cubic system is tuned by applying an external potential ($V_p$; $V_p$ = 0 corresponds to the largest inversion energy) [see the evolution of electronic structures with decreasing inversion energy for 0 GPa and 20 GPa in Supplementary Fig. S8 and Fig. S9, respectively].

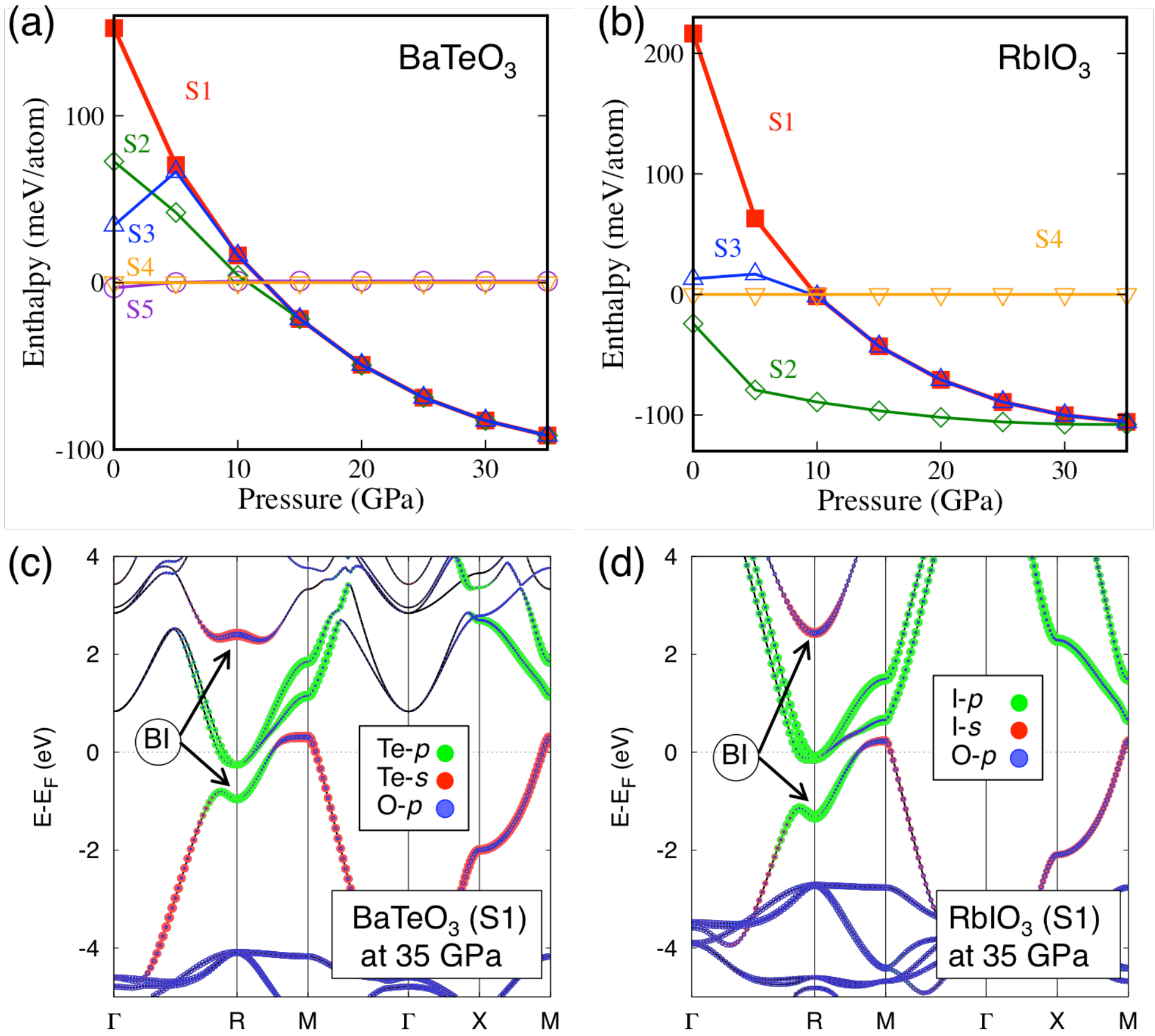

FIG. 5. Enthalpies of (a) $BaTeO_3$ and (b) $RbIO_3$ in the $ABO_3$ structures as functions of pressure, illustrating that the cubic perovskite structure (S1) with the topological gene becomes the lowest-enthalpy structure at moderately high pressure. For graphic clarity, the enthalpies of the S4 structures ($P2_1/m$) are chosen as the zero at each pressure. Swapping of A and B sites has been considered, and the lowest-energy configuration is selected. (c-d) Electronic structures of cubic $BaTeO_3$ (c) and $RbIO_3$ (d) under external pressure of 35 GPa. The band inversion is denoted in the figure by BI, with arrows pointing to the inverted states. The dotted lines with different colors denote the band projection onto different atomic orbitals.